\documentstyle[12pt,aaspp4]{article}
\tightenlines
\def\ie{i.e.\ }
\def\eg{e.g.,\ }
\def\etal{et~al.\ }

\def\ltsima{$\; \buildrel < \over \sim \;$}
\def\simlt{\lower.5ex\hbox{\ltsima}}
\def\gtsima{$\; \buildrel > \over \sim \;$}
\def\simgt{\lower.5ex\hbox{\gtsima}}

\def\msolar{${\rm M}_{\odot}$}

\newcommand{\bi}{\begin{itemize}}
\newcommand{\ei}{\end{itemize}}
\newcommand{\bc}{\begin{center}}
\newcommand{\ec}{\end{center}}

\journalid{}{}
\articleid{}{}
\lefthead{Dubinski, Mihos, \& Hernquist}
\righthead{Dark Halos and Tidal Tails}
\slugcomment{Submitted to the {\it Astrophysical Journal}}

\begin{document}

\title{Constraining Dark Halo Potentials with Tidal Tails}

\author{John Dubinski}
\affil{Canadian Institute for Theoretical Astrophysics\break
	University of Toronto, 60 St. George St.,\break
	Toronto, Ontario M5S 3H8, Canada\break
	dubinski@cita.utoronto.ca}

\author{J. Christopher Mihos}
\affil{Case Western Reserve Univ.\break
Department of Astronomy\break
10900 Euclid Ave., Cleveland, OH 44106\break
hos@burro.astr.cwru.edu}
\and

\author{Lars Hernquist}
\affil{Harvard-Smithsonian Center for Astrophysics\break
60 Garden St., Cambridge, MA 02138\break
lars@cfa.harvard.edu}

\begin{abstract}

We present an extensive parameter survey to study the influence of 
halo mass profiles on the development of tidal tails in interacting disk
galaxies.  We model the galaxies using a fixed exponential disk with a
central bulge and vary the halo potential over a range of parameters
using both the Hernquist and NFW mass distributions, probing the effect of
the halo mass, extent and concentration.  We examine the consistency of
the results against both observational and theoretical constraints on halo
profiles and comment on the failures and weaknesses of different models.
Galaxies with rising or flat rotation curves dominated by the halo
are inhibited from forming
tidal tails unless the halo is abruptly cut off just beyond the
disk edge. Conversely, models with declining rotation curves -- 
resulting either
from compact, low mass halos, or from massive disk components in
low concentration dark halos -- produce tidal tails very similar to those
observed in well-studied interacting systems.   As argued by
Springel \& White (1998), a unifying, quantitative relation 
for all cases is that the ratio of the escape velocity to the circular 
velocity at around the solar radius must be $v_e/v_c \simlt 2.5$ for tidal 
tails to be produced.  The galaxy models which appear to fit most of the 
observational constraints are those which have disk-dominated rotation
curves and low concentration halos.    We discuss our results in a
cosmological context using recent studies which link halo properties to
cosmological models.

\end{abstract}


\newpage

\section{Introduction}

Nearly 30 years have passed since the first observations of the
rotation curves of spiral galaxies revealed significant departures
from Keplerian form at large distances (\eg Rubin \& Ford 1970;
Rogstad \& Shostak 1972).  Subsequent work has verified this behavior
in large samples of spiral galaxies (\eg Rubin \etal 1980, 1982, 1985;
Kent 1987).  Unless Newtonian gravity fails on large scales or for
small accelerations, these results imply that spiral galaxies possess
massive dark halos of unseen matter (Freeman 1970).  This
interpretation is supported by estimates of the mass content of
groups and clusters of galaxies.  Furthermore, currently viable
cosmological theories invoke the existence of considerable quantities
of ``dark matter'' to account for observed properties of large scale
structure.  Indeed, a mass-to-light ratio $\sim 1000$ is required for
a critical Universe with $\Omega = 1$, far greater than those for
normal stellar populations.  (For reviews, see \eg Padmanabhan 1993;
Peebles 1993.)

While the existence of dark matter is now generally accepted, its
nature and amount remain mysterious.  For individual galaxies, these
uncertainties translate into an ignorance of the total masses and
sizes of galaxies.  In principle, these attributes could be inferred
precisely for spiral galaxies from sufficiently extended rotation
curves, but in no case has the data established Keplerian behavior
asymptotically, from which the total mass would follow.  Efforts to
measure galaxy masses using other tracers of the potential have
produced some encouraging results, most notably those based on
kinematics of satellites of the Milky Way (\eg the recent studies of
Zaritsky \etal 1989; Kochanek 1996), streams of gas and stars in the
Galactic halo (see, \eg Lin, Jones \& Klemola 1995; Johnston \etal
1998), the ``timing argument'' applied to the Local Group (\eg Kahn \&
Woltjer 1959; Peebles \etal 1989; Raychaudury \& Lynden-Bell 1989),
x-ray coronae of hot gas around ellipticals (see, \eg Forman, Jones \&
Tucker 1985), kinematics of satellites of external galaxies (Zaritsky
\etal 1993, 1997ab; Zaritsky \& White 1994), and weak gravitational
lensing of distant galaxies by closer
ones (Brainerd, Blandford \& Smail 1996).
(For reviews, see \eg Binney \& Tremaine 1987; Fich \& Tremaine 1991.)
These various methods imply that galaxies like the Milky Way are
surrounded by massive halos, but in all cases the estimates of halo
masses are uncertain to factors $\sim 2$.

Previously, it has been suggested that the tidal features associated
with {\it interacting} galaxies can be used to probe the mass
distributions in these objects.  In particular, the tidal {\it tails}
seen in a number of well-studied merging galaxies (\eg Arp 1966; Arp
\& Madore 1987) offer several advantages as tracers of the potential.
Most important, the tails in systems such as NGC 4038/39 (``The
Antennae''), NGC 7252, Arp 193, and Arp 243, to name a few, are bright
and lengthy, and hence sample the potential over a wide range in
spatial scale.  In some cases, the tails span $\sim 10 - 20$ scale
lengths of the progenitor disks in projection, likely corresponding to
$\sim 20 - 40$ scale lengths in three dimensions (\eg
Schweizer 1982; Schombert, Wallin \& Struck-Marcell 1990; Hibbard
1995).  The ``Superantennae'' is the most extreme example known, with
tails that extend $\sim 350$ kpc from tip to tip (Mirabel, Lutz \&
Maza 1991; Colina, Lipari \& Maccheto 1991).

The physics underlying the formation of tidal tails is now
well-understood, simplifying their use as a mass probe.  That
gravitational forces alone could yield such narrow structures was
illustrated by Pfleiderer \& Seidentopf (1961), Pfleiderer (1963),
Yabushita (1971), Clutton-Brock (1972ab), Toomre \& Toomre (1972),
Wright (1972), and Eneev \etal (1973).  But, it was mainly the work of
Toomre \& Toomre (1972) that emphasized the interpretation of tidal
tails as a {\it resonant} phenomenon, arising from a match between the
orbital angular frequency of the interacting galaxies and the internal
angular frequency of the disk stars.

The use of tidal features to map the potentials of interacting
galaxies was anticipated already by Faber \& Gallagher (1979), who
argued that ``Massive envelopes [halos] may well have a significant
effect on the shapes and velocities of bridges and tails created in
tidal encounters.''  Soon thereafter, White (1982) offered the
prescient remark that ``...the relative velocities of the two galaxies
at pericenter will be increased and in addition the tails will need to
climb out of a deeper potential well... Both these effects make it
more difficult to model the long tails in some observed systems.''
Indeed, as we discuss below, White's reasoning is correct, with the
shape of the halo potential being the more significant effect.
Unfortunately, the first self-consistent studies of galaxy mergers
employed a restricted set of halo parameters and were limited to
relatively low values of halo to luminous mass.  Consequently, these
models were relatively insensitive to the effects noted by White (\eg
Barnes 1988, 1992; Hernquist 1992, 1993), prompting Barnes \&
Hernquist (1992) to (incorrectly) assert, ``Since the energy
required to climb out [of potential wells] is provided by falling in,
mere length [of tails] is probably not an effective constraint on halo
masses.''

The possibility of employing tidal material to constrain properties of
dark halos was reopened by Dubinski, Mihos \& Hernquist (1996;
hereafter DMH), who performed N-body simulations of galaxy mergers
using more general galaxy models than the previous studies.  This
enabled DMH to consider a wide range of halo properties and, in
particular, to examine the influence of very massive halos on the
development of tidal tails.  These calculations demonstrated that for
the class of galaxy models adopted by DMH, the mechanisms suggested by
White (1982) do operate, and that it is indeed difficult for lengthy,
massive tails to form in galaxy collisions in which the progenitors are
surrounded by very large halos.  On this basis, DMH argued that
well-known merger candidates such as the Antennae (NGC 4038/39)
probably originated through encounters in which the progenitor
galaxies had halo to disk plus bulge mass ratios $\simlt 10:1$.  This
finding was supported by the subsequent work of Mihos, Dubinski \&
Hernquist (1998; hereafter MDH) who applied similar modeling to
photometric and kinematic data of the merger remnant NGC 7252.

At face value, these findings are at odds with expectations for the
structure of halos in cosmological models with critical mass density
(\eg CDM), $\Omega = 1$, which predict that halos at the current epoch
should be significantly more massive and extended than those favored
by the simulations of DMH (see, \eg Zurek \etal 1988; Navarro, Frenk
\& White 1996).  One interpretation of the DMH results is that their
modeling argues for a universe in which $\Omega < 1$ and in which
halos are expected to be less massive and less extended than their
counterparts in a universe with critical mass density.
Springel \& White (1998) have presented contradictory results from
simulations using dark halos motivated by $\Omega=1$ CDM
and present a possible loophole in which more extended
disks can eject long tails in massive halos.  
Clearly, our understanding of tail formation under different conditions
must be improved if we wish to use a relatively simple
argument based on the appearances of observed tidal
features to constrain $\Omega$, 
and it is important to consider the loopholes not
satisfactorily addressed by the DMH calculations.

One limitation of the DMH work was their use of a restricted set of
galaxy models for the progenitors.  Their galaxies were constructed to
provide good fits to the rotation curves of galaxies like the Milky
Way, with increasingly more mass added to the halos at large distances
to yield a sequence of models with larger ratios of halo mass to bulge
plus disk mass.  The halo profiles in these calculations were chosen
to correspond to lowered Evans models.  While a convenient choice,
such mass distributions for the halo are not motivated by cosmological
considerations.  As we demonstrate below in \S 3.1, it is the shape of
the halo potential which is mainly responsible for controlling the
lengths and masses of tails produced in mergers of disk galaxies.
Accordingly, it is desirable to consider galaxy encounters with a
wider set of halo forms than examined by DMH, to see if large halo
masses can still be accommodated by the observed properties of tidal
tails for plausible halo mass distributions.

In the modeling described here, we examine this issue by performing a
systematic survey of galaxy encounters in which we employ halos that
are motivated by cosmological simulations.  We consider two choices:
the mass distribution suggested for spherical galaxies by Hernquist
(1990), which provides a good description of halo structure when late
infall is neglected (Dubinski \& Carlberg 1991), and the NFW profile
(Navarro, Frenk \& White 1996) which has been shown to fit halos
in cosmological simulations
under a wide range of circumstances.  
The two mass models differ only in the
limit $r\rightarrow \infty$, where the NFW profile converges more
slowly than that of Hernquist (1990).  We find that tail formation is
insensitive to this difference and that our conclusions apply to
either of these two halo profiles.

As concluded previously by DMH, we again find that when we consider
galaxies designed to have flat rotation curves at small distances,
tail formation is inhibited for massive halos whose properties are
consistent with those seen in cosmological simulations of $\Omega = 1$
universes.  In such cases, the material expelled from the disks to
populate the tails falls back quickly into the merger remnant and such
models seem incapable of accounting for the tidal features seen in,
\eg the Antennae, NGC 7252, and objects with more extreme tail
attributes such as the Superantennae (Mirabel, Lutz \& Maza 1991).
Thus, the conflict noted by DMH between observed tidal tails and the
predictions of halo formation in $\Omega = 1$ cosmologies apparently
cannot be resolved by appealing to halo profiles more general than
those considered by DMH, but which are still plausible on the basis
of cosmological simulations.

In \S 2, we describe the methods employed in the simulations described
here and the galaxy models used in this study.  The outcomes of
mergers between disk galaxies with varying halo structure are described
in \S 3.  There, we demonstrate that the properties of the tails
produced in the collisions are mainly sensitive to the shape of the
halo potential, rather than differences in encounter velocity.
Observational and theoretical implications of our findings are
discussed in \S 4, and we provide conclusions in \S 5.

\section{Methods}

\subsection{A Restricted 3-body Method}

The first studies of galaxy interactions some 30 years ago used the 
restricted 3-body method to examine this problem.  Test particles 
(``stars'') were placed on circular orbits about two massive particles 
(representing each galaxy) moving on a Keplerian orbit.  This method is 
simple and computationally inexpensive yet reproduces tidal tails and 
bridges similar to those seen in much more expensive self-consistent 
N-body simulations (e.g.  Barnes 1988).  The restricted method works so 
well because the formation of tidal tails is essentially a kinematic 
effect -- stars become unbound from the disk during the interaction
and move as free test-particles in the combined potential of the
interacting galaxies.  The restricted method is ideal for extensive 
parameter surveys since it is many
orders of magnitude more computationally efficient
than are self-consistent N-body simulations.  However, if we wish to use 
the method to examine more realistic models with varying halo potentials, 
dynamical friction, and exponential disks of stars, it is necessary to 
generalize the method.

In this section, we describe a generalized, restricted 3-body method for
following the development of tidal tails during galaxy collisions.  
We retain the scheme of two moving fixed potentials but use more realistic
galaxy mass models rather than point masses.  We also move these galaxy 
potentials on a trajectory consistent with their mass distribution and 
include the loss of orbital energy through dynamical friction.  Finally, 
the phase space distribution of test-particles in the disks 
$(\vec{x}, \vec{v})$ is sampled from the distribution function of an 
exponential disk embedded within the galaxy potentials.  The test particle 
disks will therefore respond to the tidal perturbations in a similar way 
to real galactic disks.

Our generalized, restricted 3-body method works as follows:
\begin{enumerate}
	\item We assume that the galaxy potentials are rigid throughout an 
	interaction and merger.  While this is not perfectly valid, we find 
	it is adequate for examining kinematic development of tidal tails.
	\item We calculate the potential wells of the galaxies using the 
	method of Kuijken and Dubinski (1995), incorporating an exponential 
	disk, a spherical bulge and a spherical dark halo.  These models are 
	described by a 3 component distribution function and the potential 
	is calculated as the sum of a low order spherical harmonic expansion 
	and some special functions.
	\item We then calculate the expected trajectory either using a 
	previous lower-resolution simulation or an effective interaction 
	potential with a dynamical friction term discussed below.
	\item We set up a distribution of disk test-particles consistent with 
	the galaxy potential using the method of Kuijken and Dubinski (1995).
	\item We then integrate the motion of the test particles in the net
	time-dependent potential of the two rigid galaxy potentials moving 
	along the pre-calculated trajectories described above.
\end{enumerate}

This method treats the strength and duration of the tidal field during a 
collision more accurately than the point mass method and also represents the 
gradient of the outer potential realistically.
We see below that this 
generalized restricted method produces tidal tails with the same kinematic
development as those in self-consistent N-body simulations using the same
initial conditions.

\subsection{Galaxy Models}

We generate all galaxy models in our study using the method described in
Kuijken \& Dubinski (1995).  First, we use the same four ``Milky Way''
analogs (A, B, C, D) as we did in our previous studies (DMH 
and MDH).  These models have flat rotation curves within 5 exponential scale 
lengths and are approximately flat to greater distances.  We 
parameterized the halo in these 4 models in terms of a mass, 
$M = r_{c} v_c^2/G$, where $v_c$, the circular velocity, is approximately
constant out to a cutoff radius, $r_{c}$.  Beyond this radius, the
density drops off rapidly and there is very little extra halo mass.
Table 1 presents the halo:disk+bulge mass ratio along with the ratio of
escape velocity to circular velocity at the $R=2R_d$
following Springel \& White (1998).
In retrospect, our choice of mass or halo:disk+bulge mass ratio as a model 
parameter is ambiguous since the formation of tidal tails really depends on 
the shape of the potential rather than the total halo mass which can be the 
same for a variety of potentials.  We see below that the parameter
$v_e/v_c$ is a better parameter for describing the shape of the potential.
Our implicit assumption, however, is that 
all of these potentials have same shape at least out to $r_{c}$ with 
$\Phi \sim v_c^2 \log r$. While declining rotation curves may be favored on 
theoretical grounds, it is important to note that for galaxies in which
the dark matter distribution 
can be probed directly -- LSB disk galaxies -- no such 
decline is observed out to distances greater than 50 kpc (McGaugh \& de Blok 
1998).  As such, asymptotically flat rotation curves remain the most 
observationally well-motivated choice for detailed modeling.

Our original survey of four galaxy models in DMH, with the addition of a 
fifth in 
MDH, covers the behavior of tidal tail formation in only a limited manner.
Other galaxy halo models which have declining rotation curves
are certainly possible.
While such rotation curve shapes are rarely observed in nature -- and are
typically associated with the disk-halo transition in the mass distribution
rather than with the dark halo alone (\eg Casertano \& van Gorkom 1991) -- 
they are predicted by 
theoretical modeling of galaxy formation.  Dubinski \& Carlberg (1991) found 
in some cosmological simulations of galactic halo formation in the cold dark 
matter model (CDM) that the Hernquist (H) potential (1990) was a good fit to 
the halo mass distribution.  More recently, Navarro, Frenk \& White (1996) 
have found that another similar profile seems to fit halos
from many different cosmological models when late infall is included.
The H halo model is described by the potential-density pair,
\begin{equation}
	\Phi(r) = -\frac{GM}{r+r_H}
\end{equation}
and,
\begin{equation}
	\rho(r) = \frac{M_H r_H}{2 \pi} \frac{1}{r(r + r_H)^3}
\end{equation}
where $M_H$ is the total mass of the model and $r_H$ is a scale radius.
The circular velocity curve peaks at the maximum value,
\begin{equation}
	v_{H}  = 0.5 \left ( \frac{GM_H}{r_H} \right )^{1/2} \;\;\;\; {\rm at} 
\;\;\;\; r = r_H
\end{equation}
The NFW halo model is described by the potential-density pair,
\begin{equation}
	\Phi(r) = -\frac{GM_s\log(1+r/r_s)}{r}
\end{equation}
and,
\begin{equation}
	\rho(r) = \frac{M_s}{4 \pi} \frac{1}{r(r + r_s)^2}
\end{equation}
where $M_s$ is a reference mass and $r_s$ is a scale radius.
The total mass formally diverges while $M_s$ is the mass within
5.3 $r_s$.  The circular velocity peaks at the maximum value,
\begin{equation}
	v_{s} = 0.46 \left ( \frac{G M_s}{r_s} \right )^{1/2} \;\;\;\; {\rm at} 
\;\;\;\; r=2.16 r_s
\end{equation}
If we set $r_s = 0.46 r_H$ and $M_s = 0.54 M_H$ then the velocity maxima of 
the H and NFW models coincide while the NFW rotation curve drops off less 
steeply than the H rotation curve.

\subsection{Galaxy Model Parameters}

In this survey, we examine a broader range of galaxy models than in
our previous studies, with dark halos 
described by both the H model and the NFW profile.  For these models, we fix 
the disk and bulge parameters to resemble an Sa galaxy similar to 
Kuijken and Gilmore's (1989) mass model of the Milky Way.  The disk has an 
exponential surface density and an approximate ${\rm sech}^2 z$ vertical 
profile.  We use dimensionless units with the gravitational
constant $G=1$, disk mass, $M_d=1.0$, scale
length, $R_d=1.0$ and vertical scale height, $z_d = 0.15$.  The bulge is a
truncated King model with mass, $M_b=0.5$ and half mass radius, $R_b=0.3$
with a $\rho \sim r^{-3}$ density profile.  The peak circular velocity, 
$v_c$, varies according to the halo mass but $v_c \approx 1$ for most of 
the models.  For comparison to galaxies like the Milky Way a suitable choice 
of units is: $R_d = 4$ kpc, $v_c = 220$ km/s which leads to 
$M_d=4.5\times 10^{10}$ \msolar, $M_b = 2.3 \times 10^{10}$ \msolar, and a 
time unit of $18$ Myr.

With the disk and bulge parameters fixed, we set up a two parameter family 
of halos for both the H and NFW potentials.  We vary the scale radius 
($r_H$ or $r_s$) and the peak circular velocity of the halos ($v_H$ or $v_s$) 
defined above.   The range of halo scale radii is $r_H = 2.5, 5, 10, 15, 20$ 
and 25 for the H potential and a similar range of six values with 
$r_s = 0.46 r_H$ for the NFW potential chosen so the velocity peaks coincide.
We also vary the halo masses $M_H$ and $M_s$ so that the velocity peaks
cover the range $v_{s} = 0.5, 0.61, 0.71, 0.79, 0.87, 0.94$ and 1.0.
Since the mass of NFW halos is formally infinite it is necessary to 
truncate the halos at some large radius.  The cut-off is at the
radius containing a mean over-density, $\delta M/M = 18\pi^2 \approx 200$,
or the ``virial'' radius, $r_{200}$.
In cosmological simulations, $r_{200}$, as defined in NFW is approximately 15
times the scale radius $r_s$ for CDM-like halos so we set the truncation
radius at 15 $r_s$.   When the models are scaled to the Milky Way, the 
concentration parameter, $c=r_{200}/r_s$, 
formally varies for the all the models despite the fixed truncation radius  
(Table 2).  However,  we find that the physics of tail 
formation depends only on the behavior within $\sim$20 scale lengths, well 
within the truncation radius of the halos.  
In total there are 42 galaxy 
models for each of the H and NFW halo potentials.  

\begin{figure}
\centerline{\epsfbox{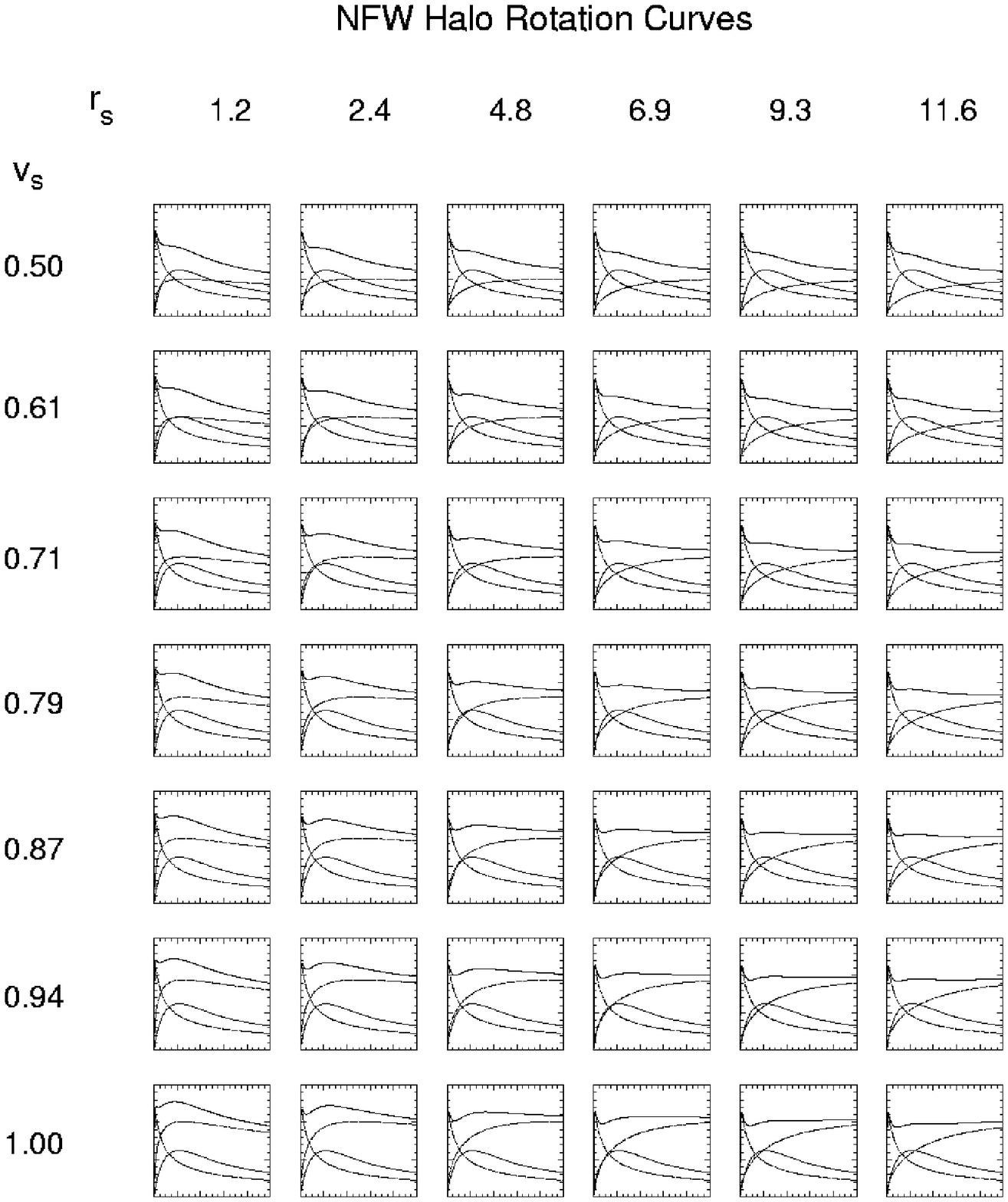}}
\caption{Rotation curve decompositions of the NFW galaxy halo models.
The net rotation curve plus contributions from the disk, bulge and halo are
shown for the 42 models - velocity scale scale is 1.5 and length scale is
10.0 in the plots.}
\label{fig-vr}
\end{figure}

The range of halo parameters is chosen to examine different aspects of
tail making.   The range of scale radii controls the halo concentration and
central density.  Halos with small scale radii have a higher central density
and therefore these
galaxies will merge more rapidly than less concentrated ones
after they collide because
of stronger dynamical friction.  Halos with large scale radii have
lower central density and impart less friction leading to longer merging
times and wider passages after the first interaction.
The range of peak circular velocities probes the effect
of the shape of the halo potential.  Smaller peak velocities correspond to
shallower potential gradients and lower escape velocities making it easier for
long tidal tails to develop.  Figure \ref{fig-vr} show the rotation curve
decompositions for the 42 NFW models.  The  rotation curves include
rising, falling and nearly flat profiles covering most of the observed range of
properties and, as we will see below, many different 
characteristics of halos from various cosmological models as well.  
Table 2 presents the formal halo
concentration parameters for models scaled to the Milky Way along with the
ratio of escape velocity to circular velocity at $R=2.0 R_d$ following
Springel \& White (1998).  The models with low escape
velocities are most likely to produce interacting pairs with long tidal tails.

\subsection{Interaction Potential}

With the galaxy models defined, we need to place them on orbital trajectories
consistent with their potentials.  Since the mass distribution is extended, 
the resulting trajectories are significantly different from Keplerian orbits.
After experimenting with different interaction potentials, we found that the
weighted average of the two galaxy potentials, 
$\Phi_1(R,z)$ and $\Phi_2(R,z)$, accurately reproduces the galaxy orbits up 
until the initial collision.  This potential is given by:
\begin{equation}
\Phi_{int}(r) = (M_2\Phi_1(R,z=0) + M_1 \Phi_2(R,z=0))/(M_1 + M_2) 
\end{equation}
where $r$ is the separation of the two galaxy centers and $M_1$ and $M_2$
are the total masses of the two galaxies.   The orbits from $\Phi_{int}$ are
equivalent to a point mass and rigid potential in orbit about their center of
mass.  Because the galaxies are centrally concentrated, this is a reasonable
approximation and the orbits very closely follow those in self-consistent, 
N-body simulations until perigalacticon.

\subsection{Dynamical Friction}
During the encounter, orbital energy is lost through dynamical friction.  
We estimate  this loss of energy (again heuristically) using the dynamical 
friction formula for a satellite sinking in an isothermal density distribution 
(Binney \& Tremaine 1987).  In the strong non-linear regime of the encounter 
of two equal mass galaxies, the Coulomb logarithm, $\ln \Lambda$, is not 
well-defined so we choose a value of $\ln \Lambda$ to fit the decay of the
orbit seen in self-consistent merger simulations.  A good compromise value 
which is a reasonable fit for galaxies with high and low mass halos is 
$\ln \Lambda = 2$.  

\subsection{Comparison to Self-consistent Interactions}

\begin{figure}
\centerline{\epsfbox{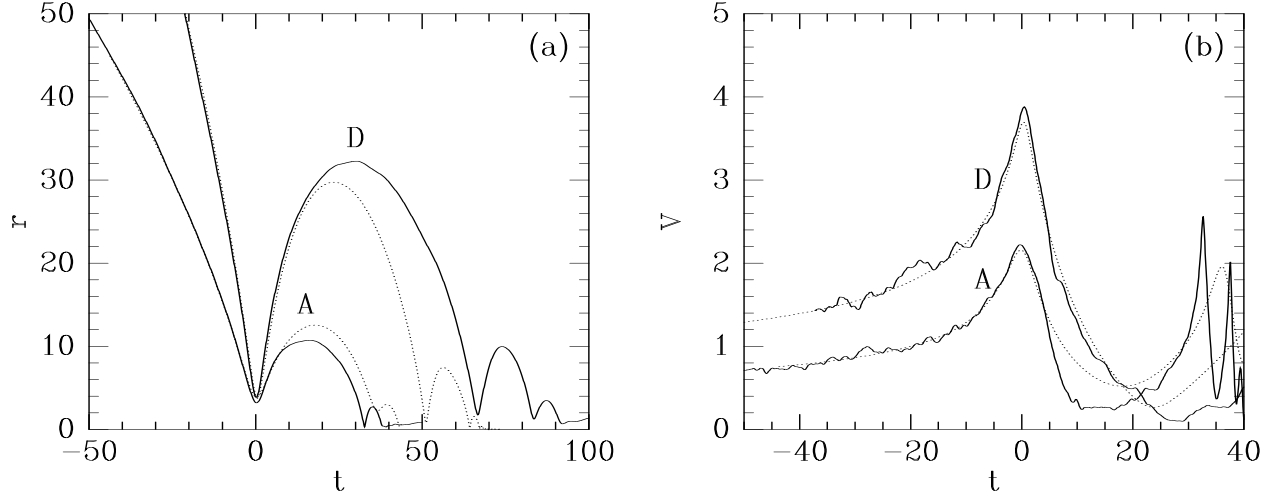}}
\caption{Comparison of galaxy orbits in a self-consistent merger and 
according to the interaction potential with a friction term. The agreement
is very good until pericenter but there is some
deviation beyond this time.  Both the
strength of the tidal perturbation at the encounter and the
merging time-scale are
well-approximated.}
\label{fig-rtvt}
\end{figure}

\begin{figure}
\centerline{\epsfbox{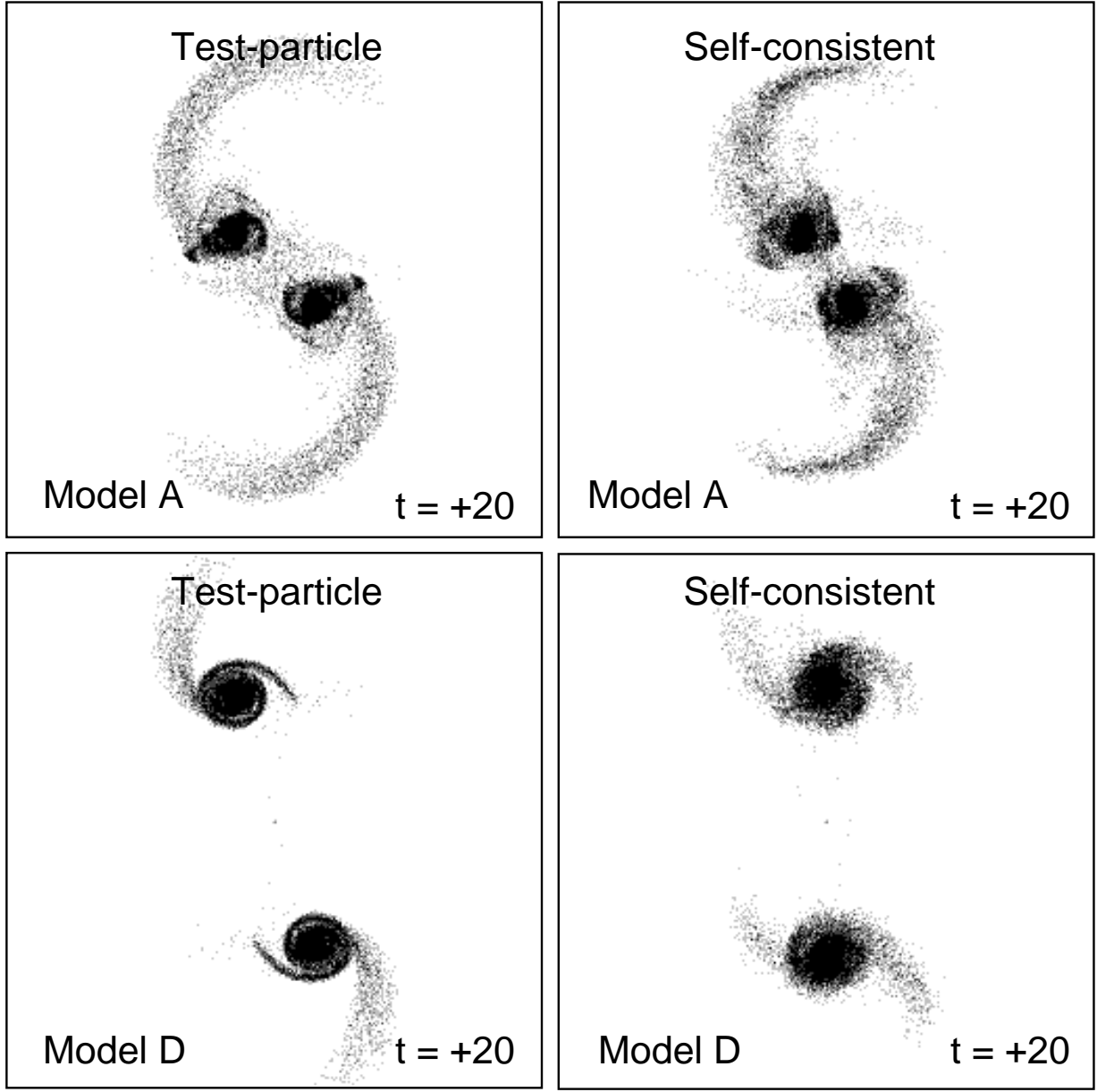}}
\caption{A comparison of the development of the self-consistent simulation
and restricted 3-body method for interacting Model A and D galaxies.
The agreement between the mass, length and kinematics of the tails is very
good.}
\label{fig-compmeth}
\end{figure}

Before applying the restricted 3-body method for our collision survey it is
useful to compare its results to those of self-consistent calculations.  
We consider two direct, co-planar equal mass galaxy collisions using 
Models A and D, a low and high mass extreme of the dark halos.  Figure 
\ref{fig-rtvt} shows the orbits calculated using the interaction potential 
with a dynamical friction term compared to the measured trajectory of the 
galaxy bulges in N-body simulations of the collision.  The orbits are almost 
identical until perigalacticon after which they are 
still relatively close but are not in exact agreement.
Since the tidal tails form after the first encounter, this treatment 
of the relative separation and speed at encounter should be very close to 
the self-consistent case.  Figure \ref{fig-compmeth} compares the appearance 
of the galaxies shortly after the interaction at the same time for both 
models.  The results are very similar, showing tidal tails with 
almost identical mass, shape, length and kinematics.  This 
good agreement emphasizes the 
point that tidal tails are a kinematic phenomenon and that the distortion of 
the halo potential at the center does not seem to have much of an effect on 
the development of the tidal tails after the interaction.

\section{Results}

\subsection{Physics of tail making}

Previous work has shown that the lengths and kinematics of tidal tails 
depend strongly on the potential of the galactic dark halo.  Colliding 
galaxies with low mass and shallow potentials yield long, massive tidal 
tails moving outwards while the high mass, steep potential counterparts 
either create short, infalling low-mass tails or inhibit tail formation 
completely.  First, the deeper potentials lead to larger relative
encounter velocities which are more impulsive and less resonant and therefore
less effective at giving disk stars the velocity perturbation they need
to be ejected into tails.  Second, the deeper and steeper potential wells of
more massive halos trap tail stars by shortening their turn-around radius 
after ejection.  Both effects combine to make the maximum length of tidal 
tails smaller in the presence of a more massive dark halo.  The observed 
{\em maximum} length of tails is therefore telling us something about the 
mass profile of the dark potential.

Two competing physical effects related to the shape of the 
potential influence the formation of tidal tails.  Both effects 
are related to the escape velocity of the galaxy potential at the radius
near the edge of the disk, where disk stars are stripped off.  First, the 
shape of the potential affects the speed of the galaxies at the time of 
collision. For zero energy orbits, the relative speed is about 
$2^{1/2} v_{esc}$.  High velocity encounters are expected in models with a 
larger halo mass and circular velocity, implying a shorter duration encounter 
and weaker tidal perturbation, resulting in shorter tidal tails.  Second, 
the gradient of the potential at the time of encounter  limits the length 
of tidal tails since steeper potentials are more effective at holding on to 
their stars.  This is a purely energetic effect since disk stars acquiring 
an energy $\delta v^2$ and in a potential with gradient, 
$\delta \Phi/\delta r$ will have length, 
$\delta r \sim \delta \Phi/\delta v^2$.  In practice, the final length and 
shape of the tails depends on the orbital and disk angular momentum along 
with the exact shape of the potential since the potential gradient itself is 
also declining with radius.  These are the two main effects governing tail 
formation but it is difficult to determine which one is the most important 
by examining the results of self-consistent simulations.

How can we disentangle these two competing effects?  One approach is to 
artificially decouple the orbital dynamics of the interacting galaxies from 
the internal dynamics of the disk stars.  In this method, galaxies are moved 
on predetermined trajectories which are not necessarily consistent with their 
mutual interaction
potential.  The trajectories are calculated first by using 
self-consistent simulations and saving the positions of the center of mass 
of the bulge.  We then move a rigid potential with its own set of test 
particles along this trajectory.  In this way, we are decoupling the effect 
of the duration of the tidal impulse from that of the shape of
the halo potential. 

We use the restricted test-particle method described above to simulate the 
formation of the tails with the following modification.  We place galaxies 
chosen from the four mass models A, B, C, and D along trajectories
derived from the self-consistent simulations of these models (the interaction 
potential gives similar results but we use real trajectories to be more 
exact). For example, we move model A galaxies along the higher speed
trajectories of model B, C, and D galaxies, and do similar experiments for 
all 16 possible permutations of halo potential and orbit.

\begin{figure}
\centerline{\epsfbox{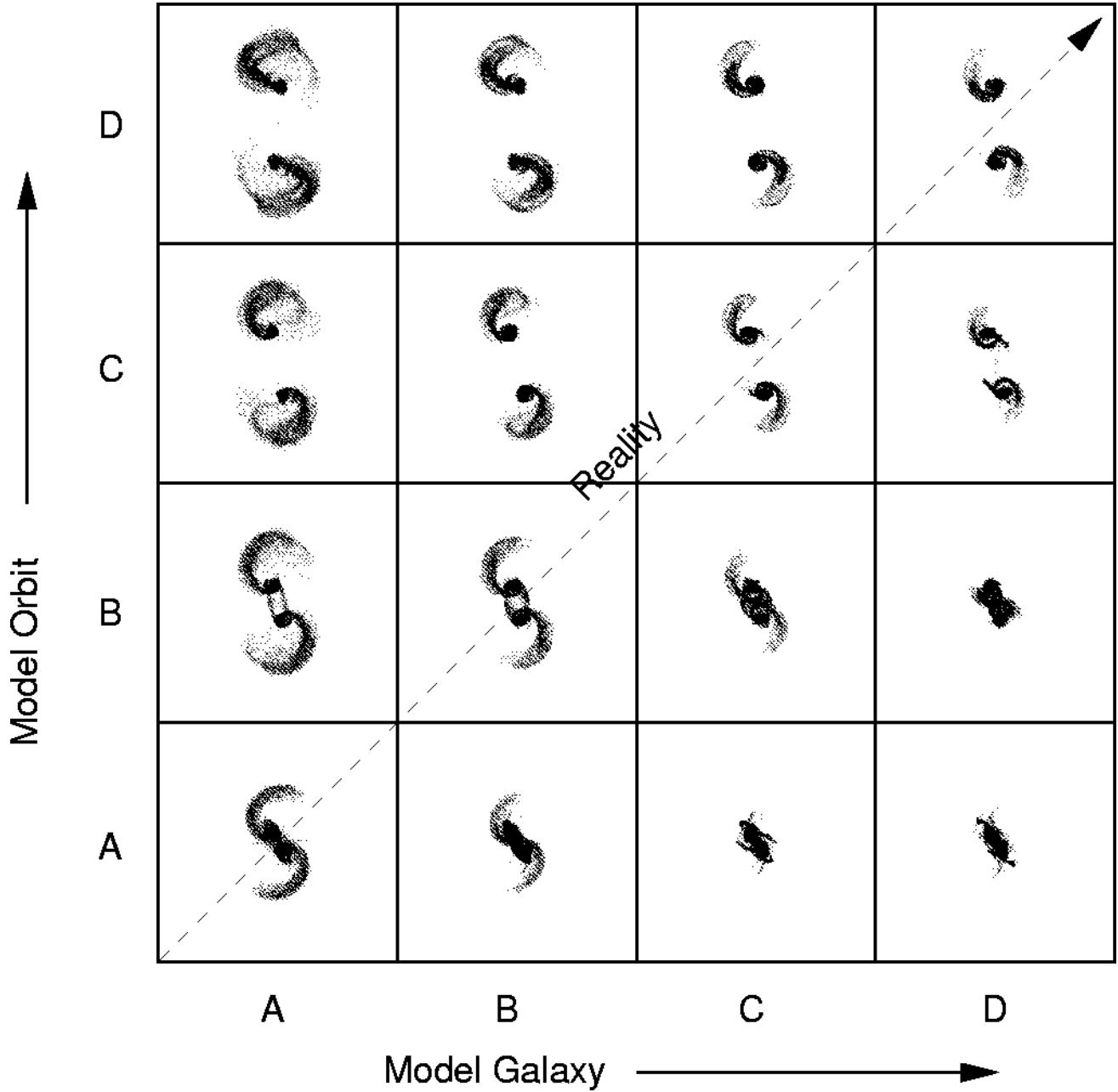}}
\caption{Discriminating the effects of the duration of the encounter from
the shape of
the potential on the evolution of tidal tails.
Galaxies models are varied along the horizontal and orbits
are varied along the vertical.  Simulations on the diagonal represent
reality while off diagonal simulations are artificial.  The upper left half
of the diagram shows that galaxies with shallow potential wells still
produce long tidal tails even when the encounters are as fast as those for
high mass galaxies.  The lower right half shows the opposite; that even in
the slow encounters of low mass galaxies the deep potential wells of high
mass galaxies still inhibit tail formation.  The implication is that the
main effect in determining a tidal tails length is the shape of
the potential well.}
\label{fig-survey}
\end{figure}

Figure \ref{fig-survey} shows the result of this exercise.  Two equal mass, 
co-planar galaxies are placed on a direct encounter with a pericentric 
distance of $R_p=4.0$.  In the figure, galaxy models are varied 
in the horizontal direction, and 
orbits are varied along the vertical.
The diagonal of this diagram corresponds to the real behavior of interacting 
pairs of galaxies from the 4 models and reveals the original effect of 
diminishing tail length and mass with halo model as discussed in DMH.  The 
off-diagonal images are artificial but reveal immediately that the 
dominant effect leading to long tidal tails is the shape of the 
potential rather than the duration of the tidal perturbation.  In the upper 
left, low mass galaxies move along the fast trajectories of high mass 
galaxies, but despite the higher speed of the encounter long tails of 
comparable length are still ejected (although on more curved trajectories).  
In contrast, the lower right of the diagram shows high mass galaxies moving 
along the low speed, rapidly merging trajectories of the lower mass galaxies.  
Despite the stronger tidal perturbation of the slow interaction, the steep 
potentials of the higher mass models still inhibit the formation of tidal 
tails.

{}From this experiment, we conclude that the dominant effect in 
determining the length and kinematics of tidal tails is simply the shape of 
the potential. The duration of the encounter does not seem to make much 
difference, since the strength of the perturbation does not vary 
significantly with 
the encounter speed.  This simplifies the interpretation of observed 
interacting galaxies since the length of tidal tails mainly reflects the 
shape of the potential, assuming that the encounters were originally within a
few scale-lengths so that the tidal perturbations were of comparable strength. 

\subsection{Varying the Halo Potentials}

With the realization that the main effect in tail formation is simply the 
shape of the potential, we carry out an extensive survey of the 
influence of 
different halo potentials on tail making using the restricted method 
described above.  We are motivated both by observational constraints on the 
properties of dark halos derived from HI and optical  rotation curves 
and cosmological predictions of the dark halo mass distribution.  For 
cosmological models, we examine halos with both the Hernquist 
and NFW forms as discussed in \S 2.2.  We find that both of these 
models give essentially the same results when scaled so that their velocity 
maxima coincide.  For brevity, we focus our discussion here on the results
obtained using the NFW profile.

\subsubsection{Direct, Co-planar Encounters}

We construct the 42 galaxy models with NFW halos described 
in \S 2.2 and simulate 
direct encounters between coplanar disks on zero energy orbits using the 
restricted method.  The pericentric 
separation of the galaxies for all trajectories is $4.0$ disk scale lengths 
(16 kpc).  Figure \ref{fig-nfw-tails} presents the results from the survey 
grid shown at $t= +30$ units (+540 Myr) after pericenter and the width of
each frame is 100 units (400 kpc).  (Animations of the simulations are 
available at {\tt http://www.cita.utoronto.ca/$\sim$dubinski/tails3}.)
The grid can be divided roughly into 3 regions which distinguish the behavior 
of the kinematic development of the tidal tails and merging times,
although there is a continuous transition between the boundaries.
We refer to these three regions as I, II and III and discuss their peculiar
features.

The first two columns of the simulation grid define Region I.  Here, galaxies 
develop long tidal tails ($>100$ kpc in length), in many ways similar to the 
classic examples of interacting pairs of galaxies discussed in TT, such as
NGC 4038/9 (the Antennae) and NGC 4676 (the Mice).  A significant amount of 
mass is also stripped from the galaxies with $\sim 20$\% of the original disk 
mass being ejected into tails.   These models have no trouble producing long, 
massive tidal tails because the halos have low mass and are compact,
resulting in a shallow potential near the disk edge.  The halo mass ranges
from only 2 to 13 times the stellar mass and the escape velocity within the
disk is only about twice the local circular velocity.  The halos at the 
bottom of this region have a higher central density and so dynamical 
friction is stronger leading to shorter merging times.  The galaxies 
in the lower left corner are already merging by $t=+30$.  The behavior in 
this region is similar to the Model A and B galaxies of DMH and, in fact, to 
most previous models of interacting galaxies.

The upper right corner of the simulation grid defines  Region II.  Here, 
galaxies also develop long tidal tails with the added bonus of a long 
connecting bridge.  The interacting pair Arp 295 with its two widely 
separated disks and bridge looks more like these models than 
those in Region I.  As in Region I, the escape velocity is again roughly 
twice the value of the circular velocity despite the higher halo mass.  
The circular speed is dropping off more rapidly at the disk edge  as well
allowing the formation of long tails and bridges.  However, the potential is
shallower than in Region I and the central halo density is much smaller and 
so the stopping power of dynamical friction is limited.  The two galaxies 
therefore fly by each other to distances greater than 50 scale lengths 
(200 kpc) before falling back together, allowing the development of a long 
connecting bridge.  The behavior is similar to the Model E galaxies 
discussed in MDH.  For $r_p=4.0$ scale lengths, the galaxies take several 
hundred time units to merge ($> 5$ Gyr) because of the smaller amount of 
dynamical friction during the first encounter, compared with merging times 
of $<1$ Gyr for galaxies in Region I.

The lower right corner of the simulation grid defines the final Region III.
Here, the formation of tidal tails is strongly suppressed, becoming 
more so as
the halo mass and extent increases towards the lower right corner.  The 
escape velocity is larger here than in Regions I and II with values ranging 
from 2.5 to 3 times the circular velocity.  The potential is therefore steeper 
and more effective at holding onto ejected material after the strong 
perturbation felt during the encounter.  The relative speed of the two 
galaxies during their encounter is 1.5 to 2 times larger than in Regions I and 
II and so the galaxies do not slow down appreciably after their encounter 
even with the heightened dynamical friction due to the higher halo densities.
The behavior seen here is similar to the Model C and D galaxies discussed in 
DMH and is primarily a result of the steep potential at the disk edge for 
these high mass halos.

In summary, the galaxies in Regions I and II produce galaxies with long
tidal tails during their first encounters with many similarities to observed 
interacting pairs.   The distribution of halo mass is markedly different in 
these regions. Models in Region I have relatively low mass, compact halos 
with declining rotation curves. In contrast, models in
Region II have higher 
mass, more 
extended halos, but their lower halo circular velocities result in 
declining rotation curves due to the disk-halo transition.  Despite these 
differences, the controlling parameter is the ratio of escape velocity to 
circular velocity within the disk ($v_e/v_c \sim 2$) rather than the actual 
halo:disk+bulge mass ratio, as emphasized by
Springel \& White (1998).  The appearance 
of the models in Region III reflects the larger amount of dark matter in the 
galaxies' centers and the steep slope of their halo potential.   It appears 
that when the escape velocity is $>2.5\times$ the circular velocity (measured 
at $2R_d$) tidal tails do not form.

\section{Discussion}

\subsection{Consistency with Observed Rotation Curves}

In contrast to our earlier studies (DMH, MDH) in which 
we used only a limited 
number of galaxy models, the expanded suite of models 
presented here provides a 
more complete map of the parameter space
of rotation curves, covering the extremes 
of observational and theoretical interpretations of the dark halo mass 
distribution.  Our models are best viewed as constraints on the depth and 
gradient of the potential well -- in other words, on the shape of the 
galactic rotation curve. We find that 
models with rotation curves which are truly flat 
inhibit the formation of tidal tails. To create long tails, rotation curves 
must decline, either through a falling halo rotation curve, or a marked 
disk-halo transition. How then do these constraints compare to other 
observational constraints on the shape of spiral galaxy rotation curves?

Optical and \ion{H}{1} studies, which at first pointed toward 
universally flat rotation curves to the limit of detectability (\eg Rubin 
\etal 1982; Kent 1987), now reveal a variety of rotation curve shapes. 
Rotation curves which are truly flat and halo-dominated -- such as 
those from isothermal 
halo models -- are extremely difficult to reconcile with observed tidal tails 
unless they have a small truncation radius (\ie just outside the edge of 
the optical/HI disk).  Models with flat rotation curves sit in the bottom 
right of Region III; these models have such steep potential gradients
in their outer regions that only very short and low-mass tails form.
The classic merging pairs of the Toomre Sequence are inconsistent with
the flat rotation curves of Region III models.

As rotation curve ``libraries'' (\eg Persic \& Salucci 1995; Mathewson \& 
Ford 1996) have expanded, however, the notion of universally flat rotation 
curves has been superseded. Instead, galaxies 
appear to possess a variety of rotation 
curve shapes, and it has been suggested that rotation curve shape is 
correlated with luminosity and/or Hubble type of the galaxy (\eg Persic, 
Salucci, \& Stel 1996; McGaugh \& de Blok 1998). Persic \etal advocate a 
``universal rotation curve'' in which bright spirals have slightly falling 
rotation curves, while low luminosity disks possess rotation curves which 
continue to rise to the last observed data point.   This classification based 
on luminosity runs into difficulties when low surface brightness (LSB) 
galaxies are considered; even luminous LSBs possess slowly rising rotation 
curves, suggesting 
that it is surface density, not luminosity, which determines 
rotation curve shape (McGaugh \& de Blok 1998).  The decline observed in some 
high surface brightness disk galaxies may in fact be associated with the 
presence of a massive disk, rather than a halo with a mass distribution 
declining faster than $r^{-2}$. As yet, no clear cut cases exist for declining 
{\it halo} rotation curves in luminous disk galaxies.

Galaxies with falling rotation curves have no problem making long tails
as shown by the examples in Regions I and II of Figure \ref{fig-nfw-tails}.  
The bottom of Region I represents models with
compact halo-dominated rotation curves that cut off at moderate radii 
(large $v_s$ and small $r_s$) while Region II represents models with 
disk-dominated rotation curves with less massive but more extended (small 
$v_s$ and large $r_s$) dark halos.  Both maximal disk models and 60\% 
maximal disk models such as advocated by Bottema (1997) are represented in 
Region II;  LSB galaxies with their extremely flat rotation curves are not.
Generally, the appearance of the tails in interacting systems such as the
Antennae and the Mice could be attributed to galaxies with either 
disk-dominated or halo-dominated rotation curves.   One possible exception 
is Arp 295 -- a widely separated pair ($> 100$ kpc) with long tails and a 
connecting bridge (Hibbard 1995).  Only disk-dominated models in Region II 
exhibit this behavior, suggesting that the galaxies in Arp 295 have 
disk-dominated rotation curves --- the galaxies in Region I simply merge 
too rapidly to allow a phase with a long connecting bridge.  The structure 
of dark matter halos probably varies from galaxy to galaxy so mergers with 
long tidal tails may simply be a product of systems with different halo 
properties but falling rotation curves.  Although galaxies with low-mass, 
compact halos (small $r_s$ and small $v_s$) represented in Region I can also 
make long tails they can be ruled out directly since their rotation curves 
simply fall too steeply.

\begin{figure}
\centerline{\epsfbox{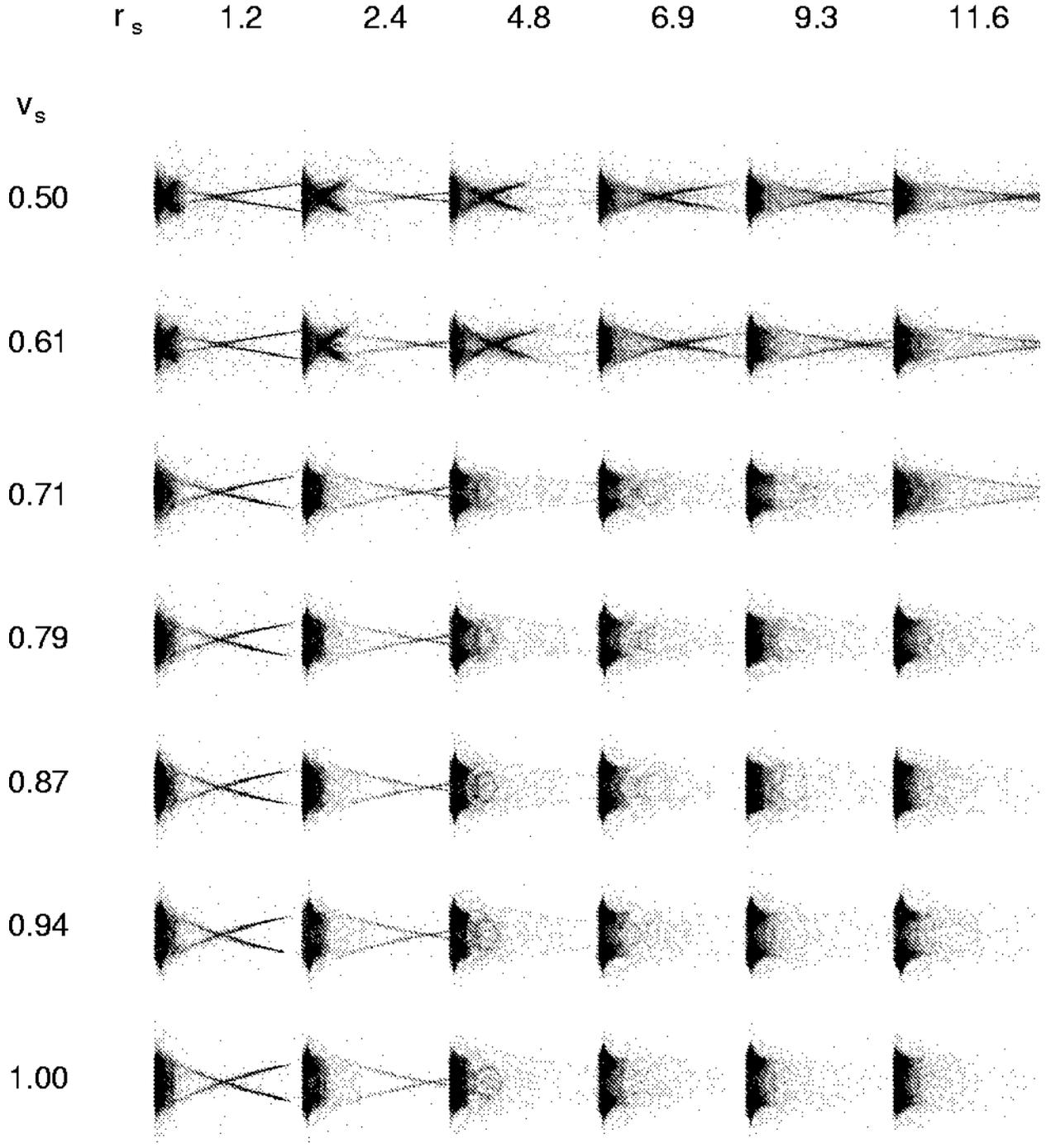}}
\caption{A survey of kinematics of the NGC 7252-like remnants.
The stellar density is plotted in the $r-v_r$ phase plane to show
the kinematical development of the tail material.  The radius is show to 30
scale lengths (120 kpc) and the velocity range is -5 to 5 (-1100 to 1100
km/s).}
\label{fig-tailsrvr7252}
\end{figure}

One further check for consistency is provided by examining late stages in
the merging process.  Figure \ref{fig-nfw-tails} provides snapshots shortly 
after an interaction as seen in systems like the Antennae or the Mice.  
NGC 7252 provides a later view when the two galaxies have coalesced with a
single, concentration of light yet memory of their interaction is still 
visible in protruding tails.  We examined a late stage analogous to NGC 7252 
by repeating the calculations using a smaller pericentric separation,
$r_p=2.0$, to decrease the merging time to less than 5 Gyr for all the models.
NGC 7252 has been caught very close to the exact time of coalescence and so 
we have run all the models to a time when the galaxy centers are 
separated only by $< 0.2$ scale length ($< 1$ kpc).  The merging times range 
from 0.3 to 5.4 Gyr after the time of closest approach (see Table 3) 
reflecting the large variation in the strength of dynamical friction during 
the interactions.  The morphology and kinematics of the remnants and any 
protruding tails are shown in Figures \ref{fig-tailsxy7252} and 
\ref{fig-tailsrvr7252}.   The galaxies in Region I still show the tails 
from the primary encounter and some forming secondary tails.  These tails 
are very long and expanding outward similar to 
those in NGC 7252. In Region II, the 
primary tails have long since escaped and the remnant is surrounded by 
diffuse debris originating from the tidal bridge formed between the galaxies.  
There are also diffuse tails formed in the second encounter which could 
potentially be identified with those of NGC 7252.  Finally, in Region III 
the formation of tails in the second encounter is suppressed as strongly as 
in the first encounter.  The tails which do form are short-lived and 
completely infalling when they reach their maximum extent of 10 scale-lengths, 
in contrast to those of NGC 7252 which are expanding outwards (Hibbard \etal 
1994).  At face value, the galaxies in Regions I and II could plausibly be 
identified with NGC 7252 at the time of merging while those in Region III are 
unlikely candidates.  However, the merger remnants of disk-dominated models of 
Region II would be surrounded by an extended cloud of diffuse stellar and 
gaseous debris.  The extended gas in NGC 7252 appears to be largely confined 
to the tails (Hibbard \etal 1994) so perhaps the preferred mass model is that 
of the compact halo models.

Other less direct observational methods exist for probing the outer rotation
curves of spiral galaxies. Zaritsky \etal (1993, 1997ab) have studied the 
velocity dispersions of satellite galaxies around luminous spirals, and find
no evidence for a drop in the velocity dispersion out to a distance of 400 
kpc. If rotation curves are truly flat to such large distances (see also 
Barcons, Lanzetta, \& Webb 1995), it is impossible to reconcile our models
with observed merging galaxies. However, the data of Zaritsky \etal do permit 
modestly falling rotation curves, for which our models do produce passable 
tidal tails (\eg the $r_s=4.8, v_s=0.71$ model).  ``Better'' tails (longer, 
more well defined) need more rapidly declining rotation curves than the 
Zaritsky \etal data allow.

Leonard \& Tremaine's (1990) analysis of high velocity stars implies an escape
velocity at the solar radius of 440-660 km/s or 2 to 3 $v_c$, straddling the 
critical value, $v_e \sim 2.5 v_c$ where tails are difficult to make.  The 
lower bound is about $v_e=440$ km/s based on the fastest star in their 
analysis. If indeed, $v_e > 550$ km/s in the solar neighborhood, the model 
study implies that the Milky Way will not throw off extended tidal tails in 
any future interaction with M31.

In summary, there are two requirements for galaxies to make realistic tidal 
tails in interactions. First, the ratio of escape to circular velocity at
$R=2R_d$ should be $v_e/v_c \simlt 2.5$. Secondly, rotation curves should be 
falling rather than flat or rising at the disk edge.  Both of these 
requirements can be met with disk-dominated models with large $r_s$ and 
small $v_s$ and a restricted set of halo-dominated models of moderate mass 
with small $r_s$ and large $v_s$.  Some interacting pairs appear to fit 
disk-dominated models better (such as Arp 295) while others prefer 
compact-halo-dominated models (NGC 7252). 

\subsection{Theoretical Constraints}

The properties of dark halos in cosmological models have been studied
extensively, enabling in principle a test of consistency with our tidal tail
arguments.  The main theoretical uncertainty is calculating how disks settle 
within these halos -- hydrodynamic simulations are still plagued with 
difficulties due to incomplete treatments of star formation and feedback, 
while semi-analytic methods 
are not well-tested.  The claim from 
DMH that the dark:luminous mass ratio of galaxies must be less than 10 was 
based on a restricted set of halo models; it is useful to expand and refine
the arguments using the simulation survey done here in order to set better
constraints. 

Most recently, Navarro, Frenk, \& White (1997) have analyzed halo formation
in a variety of cosmological models, showing how halo properties depend on 
the cosmological parameters (\eg $\Omega, \Lambda, n$).  Their main result
is that the properties of dark halos can be quantified by a
single parameter, the halo concentration, $c=r_{200}/r_s$, which is the
ratio of the ``virial radius'', $r_{200}$, to the scale-radius of the 
fitted NFW profiles.  
In general, the concentration various with mass scale in
different cosmological models with the mean value depending on the
cosmological parameters and choice of normalization of the power spectrum.
In principle, the measurement of the
concentration then can constrain a cosmological model as exemplified by
Navarro's (1998) attempts to estimate $c$ from rotation curves.

The structure and kinematics of tidal tails are similarly sensitive to 
halo concentration.   We can calculate the formal concentration for our
models by finding the effective $r_{200}$ assuming the galaxies are scaled
to galactic dimensions. The concentration parameter is
found by solving:
\begin{equation}
0.047 \left( \frac{v_s}{H_o r_s} \right)^2 = \frac{c^3}{\log(1+c) - c/1+c}
\end{equation}
The formal values of $c$ based on a galaxy with Milky Way 
dimensions (length units of 4 kpc, velocity
units of 220 km/s) and $H_o=50$ km/s/Mpc
are given in Table 2 for comparison.  (When assuming $H_o = 75$ km/s/Mpc all
concentrations are systematically lowered). 

\begin{figure}
\centerline{\epsfbox{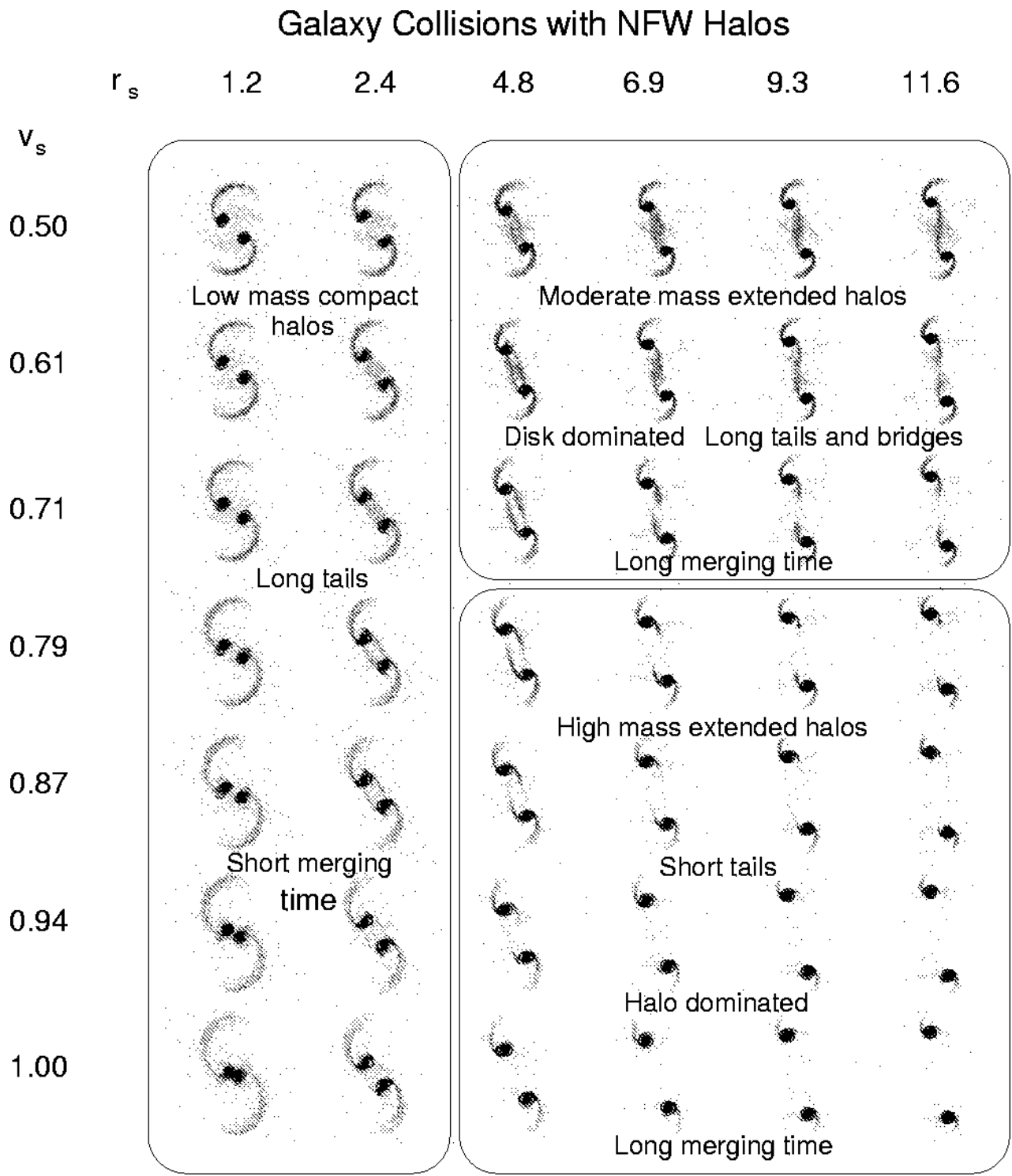}}
\caption{A survey of direct, co-planar galaxy collisions with NFW halos
having
different scale radii and maximum velocities and $R_p=4.0$.  
Each frame is 100 disk scale lengths on a side
($\sim$400 kpc).
Figure \ref{fig-vr} shows
the corresponding rotation curves of these galaxies. See the text for a
full explanation.}
\label{fig-nfw-tails}
\end{figure}

\begin{figure}
\centerline{\epsfbox{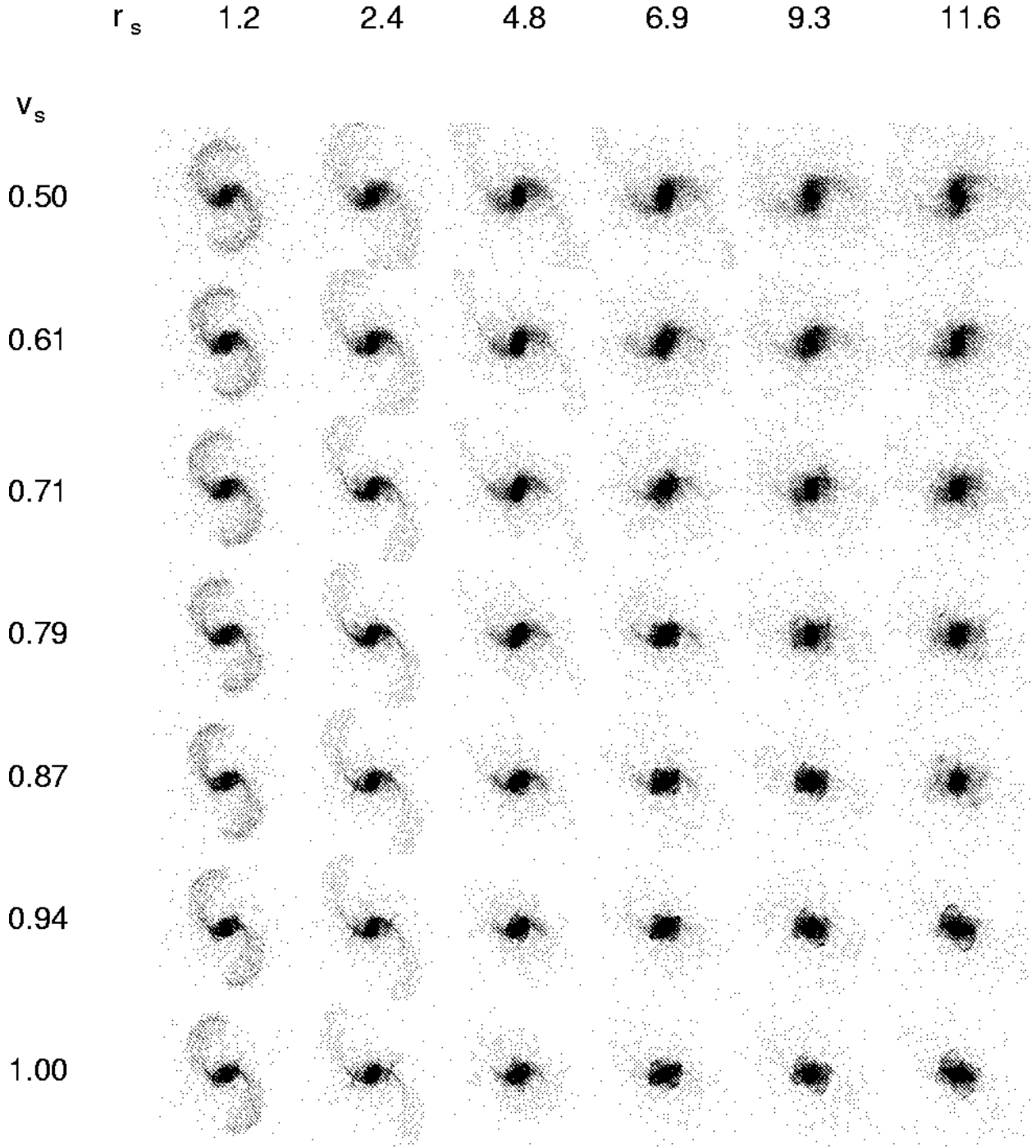}}
\caption{Same as Figure \ref{fig-nfw-tails} but
with $R_p=2.0$ and taken to the stage when the galaxies have 
just merged, as for NGC 7252.}
\label{fig-tailsxy7252}
\end{figure}

As a first comparison, we look at halos from standard CDM cosmology,
which are predicted to have $c=10-20$.
Neglecting dissipative effects on the halo, this range of 
concentrations is represented in Figure \ref{fig-nfw-tails} by the  band of 
models with $r_s=4.8$ and models with $r_s=6.9$ and $v_s > 0.8$.  
Our original $\Omega=1$ CDM-like mass models C and D lie near the bottom of the 
diagram where tails are hard to make, consistent with the original 
conclusions in DMH.  We see, however, that some models with halos having CDM 
concentrations yet which are disk-dominated in their inner regions (Region II 
models) are more forgiving. The reason for this is simply that the circular 
velocities of the disk and halo are mismatched, so that there is a decline 
in the rotation curve at the disk-halo transition.   To manufacture long 
tidal tails, in essence we must include a massive disk component to create a 
falling rotation curve within a CDM halo. These disk-dominated models are 
consistent with ``maximum disk'' models or even Bottema (1993, 1997) disk 
models, in which approximately 60\% of the total rotation curve velocity is 
attributed to the disk component. However, such models are marginally 
inconsistent with the satellite data of Zaritsky \etal (1994) for 
Milky Way-type spirals and completely incompatible with recent rotation 
curve decompositions of LSB disk galaxies (\eg de Blok \& McGaugh 1996; 
McGaugh \& de Blok 1998), which show flat rotation curves with little disk 
contribution.

In contrast to the predictions for standard CDM halos, 
the dark halos in low-density CDM models as well as flat CDM+$\Lambda$  
models are predicted to have somewhat smaller
concentrations, $c\sim 7$, when normalized to COBE (Navarro 1998).  
Models with low concentration 
halos are found at the right edge of Figure \ref{fig-nfw-tails}.
The only way such galaxies can produce long tidal tails is if the
rotation curve falls at the disk-halo transition, such as the models in
the upper right of Figure \ref{fig-nfw-tails}. 
Rotation curves which are solely supported
by low concentration halos have potential wells too steep to eject
long tidal tails.  Following the trends, we expect only disk-dominated 
models with smaller concentrations to make good tails and perhaps this 
kind of model is the most consistent theoretical choice. These
arguments suggest that the galaxies which gave rise to the Toomre Sequence
were predominantly luminous HSB galaxies with modestly declining rotation
curves due to the disk-halo transition.

We note also that when the power spectrum is normalized
instead to the number density of rich clusters in the local universe the
concentration has virtually no dependence on $\Omega$ with a constant value
of $c\approx 10$ in all cosmological models (Navarro 1998).  If this is the
correct way to normalize, then tidal tails (as well as
rotation curve shapes) cannot be used to constrain $\Omega$.  However,
the constant value of $c\approx 10$ is at least marginally consistent 
with tidal tails with this normalization.  Stronger constraints on
cosmological models must await a more thorough understanding of the
normalization of the cosmological power spectrum and how the 
formation of disk galaxies inside dark halos proceeds in 
different cosmological models.

While there is some room for consistency between our tidal tail models and the 
cosmological halo formation models of NFW, the problem of LSB galaxies still 
remains. Attempts to fit NFW models to LSB rotation curves have been made by 
McGaugh \& de Blok (1998) and Navarro (1998); while these attempts largely
failed, the ``best'' matches yielded concentration indices $c<10$, where
our models show that tidal tail formation is completely inhibited in 
halo-dominated models.  Because LSBs probably trace the dark matter component 
most faithfully, these results suggest that ``bare'' dark matter halos with 
low mass disks prohibit the formation of tidal tails; it is only the 
inclusion of a (relatively dense) baryonic component at the center to make 
a falling rotation curve necessary to produce observed tidal tails. Yet it is 
a very fine line we walk with these arguments; any significant gradient is 
inconsistent with many observed rotation curves and the satellite galaxy 
velocity dispersion measurements of Zaritsky \etal (1994).

Recently, Springel \& White (1998; hereafter SW) 
have also examined the consistency 
of long tidal tails within standard CDM halos
against observations, and it is useful to compare 
their results and interpretations to our study. SW examined disk+CDM halo 
models constructed using the prescription of Mo et al. (1998).  
According to this approach, halos are 
adjusted adiabatically in a self-consistent manner to the presence of an 
exponential disk whose length scale is determined by the initial angular 
momentum.  SW examined a
sequence of models with the same initial halo model 
($v_c \sim 250 $ km/s) but different baryon fractions and dimensionless spin
parameter, 
$\lambda$.  Their survey differs from ours in that SW fixed the halo mass 
profile and varied the disk properties, while we have fixed the disk+bulge 
mass profile and varied the halo properties.  It is encouraging that SW 
find the same range of behavior in tail making that we see in our parameter 
survey for models with similar mass profiles.  Their models cover mass 
profiles in the bottom of 
our Region I and left-hand area of Region II.  All of 
their models are halo-dominated with falling rotation curves beyond 5 scale 
lengths.  Their results emphasize that there is not necessarily a one-to-one 
correspondence between the disk scale and the halo properties, making a range 
of behaviors during galaxy interaction possible (since varying $\lambda$ 
results in a range of $R_d$). 

The models of Springel \& White span a range in disk
scale-lengths between 1.5-6.9 
$h^{-1}$.  Those models with the smallest scale lengths of 1.5-3.3 $h^{-1}$ 
kpc (their models
A,B,D) have difficulty in making long tails because disk material is
tightly bound in the inner portions of the galaxies where the model rotation
curve is not yet falling. Conversely, their models with large disk 
scale-lengths (their models C,E,F) $> 5 h^{-1}$kpc 
have no difficulty making tails --
these galaxies are endowed with a significant 
fraction of their disk material at large radii,
where the rotation curve is dropping and the potential well is shallow. 
Based on the success of the large scale-length models, Springel \& White 
claim that there is no inconsistency between observed galaxies and those 
produced in the CDM model.  While this is true in principle, it is important
to note that their successful models are large scale-length, high angular
momentum, halo-dominated systems -- the perfect description of a low
surface brightness disk galaxy. However, the rotation curves of these
models are declining, whereas LSB rotation curves remain flat out to
very large distances ($r>50$ kpc; McGaugh \& de Blok 1998). While these
models do produce long tidal tails, it is unclear whether or not they
have any real analogues in the observed galaxy population. In addition,
if only large-scale length high angular momentum systems give rise to
long tidal tails, this model would predict that the Toomre Sequence
arose predominantly from LSB progenitors. Whether or not this is
compatible with observations of the Toomre Sequence is questionable.

\section{Conclusions}

We have completed a broad survey of galaxy collisions with different mass
models to infer the processes that control the development of the
structure and kinematics of tidal tails. The main effect that governs the 
lengths and kinematics  of tidal tails is simply the shape of the potential 
well rather than strength of the perturbation during a close resonant 
encounter.  Tidal tails therefore can trace the potential in a direct way, 
relatively insensitively to the details of the encounter.  In agreement 
with Springel \& White (1998), a good criterion for making tails appears to 
be that $v_e/v_c \simlt 2.5$ at $R=2R_d$ for a wide range of models including 
those with disk-dominated and halo-dominated rotation curves.

The survey of collisions show that galaxies with rising or flat rotation 
curves effectively inhibit the formation of tidal tails.  To produce
tidal tails, galaxies must possess rotation curves which fall near the
disk edge. Two types of models meet this criteria --
galaxies with disk-dominated rotation curves and low concentration halos
(extended, moderate-mass halos) or conversely halo-dominated rotation curves
with high concentration halos (compact, low-mass halos) both create tidal
tails consistent with observed interacting galaxies. The need for a
declining rotation curve is marginally inconsistent with halo
properties inferred from the analysis of the kinematics of satellite galaxies
(Zaritsky \etal 1994) and LSB rotation curves (McGaugh \& de Blok 1998).
Observational consistency with rotation curve analyses and this galaxy 
collision study can be made if the classic tidal tail systems (\eg 
NGC 4038/39, NGC 7252, the Superantennae) arise predominantly from
luminous HSBs which are disk dominated in their inner regions and
surrounded by low-concentration dark halos. 

Galaxies embedded within standard CDM halos have high concentrations
and although these can be made to make tails if the halos are low mass, 
such halos seem to be unlikely based on rotation curve analyses.
Galaxies with massive disks embedded in low concentration halos also
produce the declining rotation curves necessary to create long tidal
tails. The fact that low concentration halos are predicted in 
cosmologies with non-zero $\Lambda$, 
coupled with recent observational developments suggesting
the presence of a cosmological constant, may argue that such disk-dominated,
low concentration halos are the most consistent model for the progenitors 
of the Toomre sequence in our universe.
Low surface brightness galaxies, with their
extended, flat, and halo-dominated rotation curves cannot make long
tidal tails due to the shapes of their potential wells. 
A strong caveat is that galaxy halo concentration is only weakly
dependent on $\Omega$ and $\Lambda$ when the power spectrum is normalized
to the local abundance of rich clusters and so tidal tails (and galaxy
rotation curves) do not necessarily provide strong cosmological
constraints.

Even with this expanded set of models, unexplored effects remain.
The models remain parameterized by only a few density profiles
(\eg the NFW and Hernquist profiles); whether or not either of
these parameterizations is correct remains unclear. Further advances 
in these comparisons will require a better understanding of disk galaxy
formation, as well as stronger observational constraints on the shapes 
of rotation curves at large radius. In addition, our models remain
focused on parabolic encounters of two distinct disk galaxies. How
mergers evolve in a cosmological setting, where the presence of a
background potential (\ie a cluster or common halo) may affect the
dynamical evolution of the pair, is still largely undetermined.
Such additional modeling would provide a useful consistency check
on our results. Once these effects are sorted out, tidal tails
may yet provide even stronger constraints on the shapes of the 
potential wells of disk galaxies.

\acknowledgments

We thank Stacy McGaugh, Volker Springel, Alar Toomre, Scott Tremaine, 
and Simon White for valuable advice.  This work was supported in part
by the NSF under Grant ASC 93--18185 and the
Presidential Faculty Fellows Program and NSERC.  JCM was supported 
by NASA through
a Hubble Fellowship grant \#~HF-01074.01-94A awarded by the Space
Telescope Science Institute, which is operated by the Association of
University for Research in Astronomy, Inc., for NASA under contract
NAS 5-26555.





\newpage

\newpage
\begin{table}
\begin{tabular}{cccccc}
&\multicolumn{5}{c}{Table 1} \\
&\multicolumn{5}{c}{Original Model H:D+B Mass Ratios} \\
&\multicolumn{5}{c}{and Escape Velocities at $R=2 R_d$} \\ \\
&A&B&C&D&E \\
\parbox{0.6in}{\bc H:D+B \\ $v_e/v_c$ \ec }&
\parbox{0.6in}{\bc  4.1  \\  2.02  \ec }&
\parbox{0.6in}{\bc  8.2  \\  2.35  \ec }&
\parbox{0.6in}{\bc  15.8 \\  2.69  \ec }&
\parbox{0.6in}{\bc  29.6 \\  3.06  \ec }&
\parbox{0.6in}{\bc  18.1 \\  2.49  \ec } \\
\end{tabular}
\end{table}

\newpage
\begin{table}
\begin{tabular}{ccccccccc}
&&&\multicolumn{6}{c}{Table 2} \\
&&&\multicolumn{6}{c}{NFW Galaxy Model Concentrations} \\
&&&\multicolumn{6}{c}{and Escape Velocities at $R=2R_d$} \\ \\
&$r_s$&1.2 & 2.4 & 4.8 & 6.9 & 9.3 & 11.6 \\
$v_{s}$&&&&&&&\\
0.50&\parbox{0.5in}{\bc $c$ \\ $v_e/v_c$  \ec}&
\parbox{0.5in}{\bc  29 \\   1.80   \ec }&
\parbox{0.5in}{\bc  17 \\   1.93   \ec }&
\parbox{0.5in}{\bc  10 \\   2.10   \ec }&
\parbox{0.5in}{\bc   7.1 \\   2.19   \ec }&
\parbox{0.5in}{\bc  5.5 \\   2.24   \ec }&
\parbox{0.5in}{\bc  4.6 \\   2.28   \ec }
\\
0.61&\parbox{0.5in}{\bc $c$ \\ $v_e/v_c$ \ec}&
\parbox{0.5in}{\bc  34 \\   1.85   \ec }&
\parbox{0.5in}{\bc  20 \\   2.05   \ec }&
\parbox{0.5in}{\bc  11 \\   2.24   \ec }&
\parbox{0.5in}{\bc  8.4 \\   2.38   \ec }&
\parbox{0.5in}{\bc  6.6 \\   2.45   \ec }&
\parbox{0.5in}{\bc  5.4 \\   2.52   \ec }
\\
0.71&\parbox{0.5in}{\bc $c$ \\ $v_e/v_c$  \ec}&
\parbox{0.5in}{\bc  38 \\   1.89   \ec }&
\parbox{0.5in}{\bc  22 \\   2.10   \ec }&
\parbox{0.5in}{\bc  13 \\   2.37   \ec }&
\parbox{0.5in}{\bc  10 \\   2.53   \ec }&
\parbox{0.5in}{\bc  7.5 \\   2.65   \ec }&
\parbox{0.5in}{\bc  6.2 \\   2.70   \ec }
\\
0.79&\parbox{0.5in}{\bc $c$ \\ $v_e/v_c$ \ec}&
\parbox{0.5in}{\bc  41 \\   1.93   \ec }&
\parbox{0.5in}{\bc  24 \\   2.18   \ec }&
\parbox{0.5in}{\bc  14 \\   2.46   \ec }&
\parbox{0.5in}{\bc  10 \\   2.66   \ec }&
\parbox{0.5in}{\bc   8.2 \\   2.78 \ec }&
\parbox{0.5in}{\bc  6.8 \\   2.87  \ec }
\\
0.87&\parbox{0.5in}{\bc $c$ \\ $v_e/v_c$ \ec}&
\parbox{0.5in}{\bc  44 \\   1.95   \ec }&
\parbox{0.5in}{\bc  26 \\   2.23   \ec }&
\parbox{0.5in}{\bc  15 \\   2.54   \ec }&
\parbox{0.5in}{\bc  11 \\   2.77   \ec }&
\parbox{0.5in}{\bc  8.8 \\   2.89  \ec }&
\parbox{0.5in}{\bc  7.4 \\   3.01  \ec }
\\
0.94&\parbox{0.5in}{\bc $c$ \\ $v_e/v_c$ \ec}&
\parbox{0.5in}{\bc  47 \\   1.98  \ec }&
\parbox{0.5in}{\bc  28 \\   2.26  \ec }&
\parbox{0.5in}{\bc  16 \\   2.63  \ec }&
\parbox{0.5in}{\bc  12 \\   2.86  \ec }&
\parbox{0.5in}{\bc  9.4 \\  3.02 \ec }&
\parbox{0.5in}{\bc  7.9 \\  3.14 \ec }
\\
1.00&\parbox{0.5in}{\bc $c$ \\ $v_e/v_c$ \ec}&
\parbox{0.5in}{\bc  49 \\   1.99  \ec }&
\parbox{0.5in}{\bc  29 \\   2.30  \ec }&
\parbox{0.5in}{\bc  17 \\   2.68  \ec }&
\parbox{0.5in}{\bc  13 \\   2.93  \ec }&
\parbox{0.5in}{\bc  10 \\   3.09  \ec }&
\parbox{0.5in}{\bc  8.3 \\  3.22 \ec }
\\
\end{tabular}
\end{table}

\newpage
\begin{table}
\begin{tabular}{ccccccccc}
&&\multicolumn{7}{c}{Table 3} \\
&&\multicolumn{7}{c}{Merging Times in Gyr for $r_p=2.0$} \\ \\
&$r_s$&1.2 & 2.4 & 4.8 & 6.9 & 9.3 & 11.6 \\

$v_{s}$&&&&&&& \\

0.50 & &   0.52& 0.86& 1.73& 2.84& 4.13& 5.57 \\
0.61&&     0.49& 0.85& 1.78& 2.91& 4.24& 5.71 \\
0.71&&     0.47& 0.81& 1.71& 2.78& 4.03& 5.47 \\
0.79&&     0.44& 0.75& 1.63& 2.65& 3.83& 5.14 \\
0.87&&     0.41& 0.74& 1.55& 2.50& 3.59& 4.81 \\
0.94 &&    0.39& 0.69& 1.49& 2.37& 3.42& 4.39 \\
1.00&&     0.39& 0.68& 1.42& 2.26& 3.13& 4.33
\end{tabular}
\end{table}


\end{document}